\documentclass[11pt]{article}
\usepackage[utf8]{inputenc}
\usepackage[T1]{fontenc}
\usepackage{lmodern}
\usepackage[margin=1.1in]{geometry}
\usepackage{setspace}
\usepackage{microtype}
\usepackage{booktabs,tabularx,array}
\usepackage{ragged2e}
\usepackage{xurl}
\usepackage[colorlinks=false,hidelinks]{hyperref}
\usepackage{authblk}
\usepackage[backend=biber,style=numeric,sorting=none,maxbibnames=3,minbibnames=3,giveninits=true,natbib=true,doi=true,url=true,isbn=false]{biblatex}
\DeclareNameAlias{default}{family-given}

\title{When AI Performance Is Not Enough: Statistical Guardrails for Clinical Development}
\author[1]{Siddhesh Kulkarni}
\author[2]{Gina D'Angelo}
\author[3]{Swarna Reddy}
\author[4]{Rakhi Kilaru\thanks{Corresponding author: rakhi.kilaru@thermofisher.com}}

\affil[1]{Bristol Myers Squibb, 5 Giralda Farms, Madison, NJ 07940, USA}
\affil[2]{Statistical Innovation, Oncology R\&D, AstraZeneca, Gaithersburg, MD 20878, USA}
\affil[3]{ASA AIML Clinical Sub-Team, USA}
\affil[4]{Thermo Fisher Scientific, 3900 Paramount Parkway, Morrisville, NC 27560-7200, USA}
\date{}
\begin{document}
\maketitle
\begin{abstract}
\noindent
Artificial intelligence (AI) and machine learning (ML) are reshaping the
operational and inferential landscape of clinical development, yet their
adoption frequently outpaces the statistical scrutiny these methods demand.
This commentary offers first-principles perspectives across four application
domains: anomaly detection in clinical trial data, real-time monitoring and adaptive trial designs, biomarker-based enrichment, and virtual and digital twins. 
For each domain, we articulate the underlying inferential problem, the contribution of AI and ML relative to established statistical methods, and the risks that statisticians, data scientists, and clinical developers must actively manage. 
Across these domains, a unifying theme emerges: AI and ML methods may succeed locally while failing to satisfy the broader statistical assumptions required for valid clinical and regulatory decisions. These risks include false-positive signals, compromised Type~I error control in adaptive designs, conflation of prognostic with predictive biomarkers, and causal claims not supported by data and assumptions.
We argue that a clearly defined context of use, relevant statistical assumptions, and consequences of the intended decision should guide the evaluation. For treatment-effect questions, the ICH~E9(R1) estimand framework provides a useful organizing principle. Responsible use of AI in clinical development requires not restraint, but rigor: the inferential discipline necessary to make these technologies reproducible, interpretable, and defensible in the service of patients.
\end{abstract}

\section{Introduction}
The question facing clinical development is no longer whether AI and ML
will be used but how their use can be made statistically coherent and
regulatorily defensible. These methods are not a departure from statistical
thinking; in many cases, they are extensions of it, often more flexible,
occasionally more opaque, and always in need of the inferential scaffolding
that statisticians are trained to provide. Strong predictive performance
alone does not establish the validity of clinical and regulatory decisions;
this requires consideration of the intended use, statistical assumptions,
and consequences of acting on model outputs.

This commentary considers four domains with substantial opportunity and
under-appreciated risk: anomaly detection in clinical trial data, real-time monitoring and adaptive trial designs, biomarker-based enrichment, virtual and digital twins.
The integration of multimodal data is discussed within these applications
as relevant. Each section addresses the statistical problem, the contribution
of AI and ML, and perspectives on responsible implementation.

\section{Anomaly Detection in Clinical Trial Data}
Anomaly detection in clinical trials spans a methodological continuum, and
the appropriate approach is dictated less by algorithmic sophistication than
by the detection problem: what is being flagged, on what unit of analysis,
relative to what reference, and what decision the output will inform.
\emph{Traditional statistical methods}, including centralized statistical
monitoring, provide interpretable and auditable signals for data-quality
and site-level monitoring \citep{venet2012,trotta2019}.
Machine \emph{learning methods}, such as random forests and gradient-boosted
trees, can learn nonlinear relationships from tabular data, while
\emph{deep learning methods} can extract representations from unstructured
and multimodal input. Greater model complexity does not guarantee better
performance; general tabular benchmarks show that tree-based methods remain
competitive \citep{grinsztajn2022,shwartzziv2022}.

Across these approaches, an unusual pattern is a signal for investigation,
not evidence of erroneous data or misconduct. Site differences may reflect
patient characteristics, measurement practices, or random variation, while
repeated screening can generate false alerts \citep{trotta2019}. Research has
also evaluated changes in site-level data-quality metrics following monitoring
\citep{deviron2024}. Validation should therefore consider sensitivity,
false-alert burden, cross-site comparability, and follow-up consequences.
Consistent with risk-based regulatory thinking, credibility requirements
should reflect context of use and decision consequence rather than the
``AI'' label alone \citep{fda2025ai_draft,ema2024ai}.

\section{Real-time Monitoring and Adaptive Trial Designs}
AI-enabled real-time monitoring has been explored by Takeda in partnership
with MIT \citep{lin2022BioPharm}. Their Statistical Monitoring in Real Time
(SMRT) tool uses probabilistic programming to build generative models from
accumulating trial data and identify unusual patterns in eligibility,
safety, and efficacy data. These signals can support timely investigation
of potential data-quality issues.

In April 2026, the FDA announced its real-time clinical trial (RTCT)
initiative, allowing agency scientists to review predefined safety signals
and endpoints as trials progress \citep{fda2026newsrelease}. AstraZeneca's
TRAVERSE and Amgen's STREAM-SCLC studies were initiated as proofs of concept. 
FDA reported successful receipt and validation of signals from TRAVERSE,
demonstrating technical feasibility of real-time signal sharing. This
highlights sponsor--regulator agreement on signal criteria, reliable data
integration, and validation of signals.

Importantly, real-time monitoring does not itself constitute an adaptive
trial design. Adaptive designs involve prospectively planned modifications
based on interim analyses such as early stopping, sample-size re-estimation,
or response adaptive randomization. AI and ML may support timely processing and
decision-making, but their integration must preserve trial integrity and
valid inference. Prespecified adaptation rules, operational safeguards,
and simulation-based evaluation of Type~I error, power, and treatment-effect
estimation remain essential \citep{fda2019adaptive,ich2025e20draft}.
Timely and reliable interim analyses are also necessary to realize the benefits
of adaptive designs \citep{thomson2026interim}.

\section{Biomarker-Based Enrichment}
Precision medicine has increased the incorporation of biomarkers into clinical
trials, supporting targeted therapies for biologically defined populations.
Regulatory frameworks recognize the importance of linking a therapeutic
product with a validated biomarker strategy, including companion diagnostics
when testing is essential for safe and effective use
\citep{fda2014invitro,fda2016CDx}. For new compounds, human data may be
limited, and teams must identify relevant biomarkers, assays, and cutoffs
while balancing speed with scientific validity \citep{fda2019enrichment}.

AI can integrate genomic, proteomic, imaging, electronic health record, and
clinical-trial data to identify biomarkers associated with treatment response
\citep{Vamathevan2019}. Its usefulness nevertheless depends on data quality,
representativeness, and provenance. Differences in measurement, missingness,
and assay platforms can introduce bias and batch effects; associations from
one population may not generalize to another.

Crucially, AI-generated associations do not necessarily demonstrate that a
biomarker predicts treatment benefit. A biomarker may be prognostic,
reflecting disease outcome regardless of treatment, rather than predictive,
identifying patients who derive differential benefit from a therapy.
Distinguishing these concepts requires appropriate treatment comparisons,
evaluation of treatment--biomarker interactions, and attention to multiplicity
\citep{Ballman2015}. Analyses of randomized trials illustrate that ML
estimates of heterogeneous treatment effects may fail to validate
independently \citep{chen2026hte}.

Accordingly, AI-supported enrichment requires a defined context of use,
prespecified analysis, analytically reliable assays, and independent
validation \citep{fda2025ai_draft}. AI-derived findings should remain
hypotheses until supported by appropriate clinical and regulatory evidence.

\section{Virtual and Digital Twins}
Digital twins are emerging as an ambitious application of AI and ML in
clinical development because they move beyond static risk prediction
toward patient-specific or population-specific simulation. By adjusting for prognostic scores and integrating
longitudinal and multimodal data, including biomarkers, imaging, and
electronic health records, they may support disease-progression modeling,
patient enrichment, and synthetic or hybrid control arms
\citep{pappalardo2019,wang2022insilico}.

The central statistical challenge of implementing virtual twins, however, is not whether a model can
generate plausible trajectories, but whether those trajectories are credible
for the decision they are intended to inform. Predicting future outcomes
under observed treatment addresses prediction; projecting outcomes under an
alternative treatment makes a counterfactual claim. Interpreting differences
as treatment effects requires a defined estimand and defensible causal
assumptions \citep{iche9r1,ruhrbergestevez2026}.

Exploratory prediction requires calibration and external validation. Synthetic or hybrid controls additionally raise concerns about exchangeability,
endpoint comparability, missing data, and transportability
\citep{fda2023externalcontrol}. Longitudinal treatment changes and biomarkers
may create treatment--confounder feedback that ordinary regression cannot
adequately address \citep{robins2000}. Higher-consequence applications require
prespecified credibility plans, uncertainty assessment, and sensitivity
analyses, with rigor matched to context of use
\citep{fda2025ai_draft,cicin2026digitaltwins}.

The value of digital and virtual twins is not that they replace randomized evidence, but
that they organize patient-level information into testable, decision-relevant
simulations while identifying where prospective evidence is needed.

\section{Discussion}
Across the four domains, a recurring challenge is the gap between algorithmic
performance and statistical credibility for an intended clinical-development
decision. An anomaly flag need not indicate a data-quality failure; real-time
monitoring does not establish adaptive-design validity; a biomarker associated
with outcomes need not predict treatment benefit; and a plausible digital-twin
trajectory does not establish a causal effect.

We propose a common decision-centered approach: first, define the context of
use and intended decision; second, specify the statistical target or estimand,
where appropriate, and its assumptions; third, assess performance, uncertainty,
and potential bias; and finally, establish validation and sensitivity analyses
proportionate to the decision's consequences. Multimodal data integration
requires additional attention to measurement consistency, missingness, and
transportability. This approach aligns with risk-based AI credibility principles
and, for treatment effects, ICH~E9(R1)
\citep{fda2025ai_draft,ema2024ai,iche9r1}.
Table~\ref{tab:decision} summarizes how these requirements differ by application.

\section{Conclusions}
Responsible use of AI and ML in clinical development requires more than
accurate predictions or plausible simulations. It requires statistically
defensible evidence that model outputs can reliably support the clinical and
regulatory decisions for which they are intended.

\section*{Acknowledgments}
We acknowledge ASA Biopharmaceutical section's AI/ML Clinical Sub-team working group. 
\section*{Funding}
None to declare.
\section*{Declaration of Competing Interests}
Authors declare no competing interests. 
\section*{Author Contributions}
All authors contributed equally.
\clearpage
\printbibliography[title={References}]
\clearpage
\begin{table}[p]
\centering
\caption{From AI/ML outputs to statistically credible decisions in clinical development.}
\label{tab:decision}
\setstretch{1.2}
\small
\begin{tabularx}{\textwidth}{@{}>{\RaggedRight\arraybackslash}p{2.8cm}>{\RaggedRight\arraybackslash}p{3.2cm}>{\RaggedRight\arraybackslash}p{3.65cm}>{\RaggedRight\arraybackslash}X@{}}
\toprule
\textbf{Domain} & \textbf{Intended decision} & \textbf{Statistical failure to avoid} & \textbf{Decision-specific evidence} \\
\midrule
Anomaly detection & Investigate atypical site patterns & False alerts or mistaking atypicality for misconduct & Contextual review; sensitivity and false-alert burden \\
\addlinespace
Real-time monitoring and adaptive designs & Modify trial conduct at interim analyses & Invalid operating characteristics after adaptation & Prespecified rules; simulated Type~I error, power and bias \\
\addlinespace
Biomarker enrichment & Select a treatment-responsive subgroup & Confusing prognosis with differential treatment benefit & Treatment interaction; multiplicity control; independent validation \\
\addlinespace
Virtual and digital twins & Use simulated patient trajectories & Unsupported causal or transportability claims & Calibration; identification assumptions; sensitivity analyses \\
\bottomrule
\end{tabularx}
\vspace{1.0em}
\begin{minipage}{\textwidth}\footnotesize
\textit{Note:} The proposed framework summarizes decision-specific statistical
considerations; it is not a newly validated regulatory checklist.
\end{minipage}
\end{table}
\end{document}